\documentclass[a4paper,11pt]{article}
\usepackage{pos}

\title{HAL QCD potentials with non-zero total momentum and an application to the $I=2$ $\pi\pi$ scattering}
\ShortTitle{HAL QCD potentials with non-zero total momentum}

\author*[a,b]{Sinya Aoki}
\author[a,b]{Yutaro Akahoshi}

\affiliation[a]{Center for Gravitational Physics, Yukawa Institute for Theoretical Physics, Kyoto University,\\
Kitashirakawa Oiwakecho, Sakyo-ku, Kyoto 606-8502, Japan}

\affiliation[b]{Riken Nishina Center (RNC),\\
Hirosawa 2-1, Wako, Saitama 351-0198, Japan}

\emailAdd{saoki@yukawa.kyoto-u.ac.jp}
\emailAdd{yutaro.akahoshi@yukawa.kyoto-u.ac.jp}

\abstract{
We consider the HAL QCD method in the system with non-zero total momentum (laboratory frame).
We derive a relation between the NBS wave function in the laboratory frame and the energy-independent non-local potential (HAL QCD potential), and propose the time-dependent method to extract the potential from correlation functions in the laboratory frame.
We then apply this formulation to the $I=2$ $\pi\pi$ system to calculate the corresponding potential in the laboratory frame, employing 
the 2+1 flavor gauge configuration on a $32^3\times 64$ lattice at the lattice spacing $a\simeq 0.091$ fm and $m_\pi \simeq 700$ MeV.
While statistical errors are larger, the effective leading order (LO) potentials and corresponding phase shift agree with those
from the HAL QCD potential in the center of mass (CM) frame.
We also demonstrate the consistency in scattering phase shifts between the HAL QCD method in several frames and the finite volume method. 
The HAL QCD method in the  laboratory frame enlarges applicabilities of the method to investigate hadron interaction including  
mesonic resonances such as $\rho$ and $\sigma$.
}

\FullConference{%
 The 38th International Symposium on Lattice Field Theory, LATTICE2021
  26th-30th July, 2021
  Zoom/Gather@Massachusetts Institute of Technology
}

\begin{document}
\begin{flushright}
YITP-21-139
\end{flushright}
\maketitle

\section{Introduction}
Hadron interactions have been investigated actively in lattice QCD mainly by two methods,
the finite volume method\cite{Luscher:1990ux,Rummukainen:1995vs,Hansen:2012tf} and the HAL QCD method\cite{Ishii:2006ec,Aoki:2009ji,Aoki:2011ep,HALQCD:2012aa},
both of which are based on the fact that the Nambu-Bethe-Salpeter (NBS) wave functions contains information of the scattering S-matrix in QCD.
 
 The HAL QCD method works well, in particular for two baryon interactions (see \cite{Aoki:2020bew} and references therein),
 thanks to the time-dependent method\cite{HALQCD:2012aa} and multi-channel extension\cite{Aoki:2011gt}.
Recently, resonances such as the $\rho$ meson have been investigated in the HAL QCD method,
by various improvements for all-to-all propagators\cite{Kawai:2017goq,Kawai:2018hem,Akahoshi:2019klc,Akahoshi:2020ojo,Akahoshi:2021sxc}. 
If we extend the HAL QCD method to systems having the same quantum numbers of the QCD vacuum such as the $\sigma$ meson,
a serious obstruction appears.  For example, 
the vacuum contribution dominates  over signals for $\pi\pi$ states in
the correlation function between $\sigma$ resonance and the $I=0$ S-wave $\pi\pi$ 
in the center of mass frame as
\begin{equation}
\langle 0\vert \pi(t)\pi(t) \sigma(0) \vert 0\rangle \simeq \langle 0\vert \pi(t)\pi(t)\vert 0\rangle\langle 0\vert \sigma(0) \vert 0\rangle
+ e^{-2m_{\pi} t } \langle 0\vert \pi(t)\pi(t)\vert \pi\pi\rangle\langle \pi\pi\vert \sigma(0) \vert 0\rangle+\cdots.
\end{equation}
An introduction of  non-zero total momenta to the system is a promising way to remove this type of contaminations from the vacuum, and the theoretical formulation for the HAL QCD method in the laboratory system has already been proposed\cite{Aoki:2019pnq}.
Recently, we have performed a numerical test of the formulation to extract the potential from the $I=2$ $\pi\pi$ system in the laboratory frame, which is reported here. After briefly explaining the general formulation, we present numerical results.
The $I=2$ $\pi\pi$ scattering phase shift obtained in the laboratory frame are compared with those from the potential in the CM frame as well as the finite volume spectra. 

\section{ Formulation for  the HAL QCD potential in the laboratory frame}
\subsection{Lorentz transformation for the NBS wave function}
 The NBS wave function for a scalar theory in the Minkowski spacetime is defined in  a general frame as
 \begin{equation}
 \psi_{k_1,k_2}(x_1,x_2) =\langle 0\vert T\left\{\phi(x_1)\phi(x_2)\right\}\vert k_1,k_2\rangle
 :=\varphi_{k_1,k_2}(x) e^{-i W X^0 + i {\bf P}\cdot{\bf X}},
 \end{equation}
 where $\phi(x)$ is a scalar field operator, $\vert k_1,k_2\rangle$ is an asymptotic in-state of two particles with four momenta $k_1$ and $k_2$, $W:=\sqrt{{\bf k}_1^2+m^2} + \sqrt{{\bf k}_2^2+m^2}$ and ${\bf P}:={\bf k_1}+{\bf k}_2$ are the total energy and momentum, respectively, while $X:=(x_1+x_2)/2$ and $x:=x_1-x_2$ are center of gravity and relative coordinates, respectively, and $\varphi_{k_1,k_2}(x)$ is a relative NBS wave function.
 
 The relative  NBS wave function in the laboratory frame is related to the one in the CM frame as
 $\varphi_{k_1,k_2}(x) = \varphi_{k_1^*,k_2^*}(x^*)$, where quantities with $*$ represent those in the CM frame, and
 are related to those without it in the laboratory frame as
$x^{* 0} =\gamma (x^0-{\bf v}\cdot {\bf x}_{\parallel})$, $ {\bf x}^*_\parallel = \gamma({\bf x}_\parallel -{\bf v} x^0)$, 
${\bf x}^*_\perp ={\bf x}_\perp$. Here $\parallel$ and $\perp$ means vectors parallel and perpendicular to ${\bf v}:= {\bf P}/W$, respectively.  The boost factor $\gamma$ is given by $\gamma:={1\over \sqrt{1-{\bf v}^2}}$,   $W^{*2}=W^2-{\bf P}^2$, and ${\bf P}^*=0$ by definition.

\subsection{Potential from the NBS wave function in the laboratory frame}
The HAL QCD potential is defined in the CM frame, and is extracted from the correlation functions in the Euclidean spacetime by the Wick rotation as $x^4=ix^0$, which leads to
$x^{* 4} =\gamma (x^4- i{\bf v}\cdot {\bf x}_{\parallel})$, $ {\bf x}^*_\parallel = \gamma({\bf x}_\parallel + i{\bf v} x^4)$, 
${\bf x}^*_\perp ={\bf x}_\perp$. 

The NBS wave function is related to the non-local potential in the derivative expansion as
\begin{eqnarray}
{1\over 2\mu} (\nabla^{*2}+{\bf k}^{*2} ) \varphi_{k_1^*,k_2^*}({\bf x}^*, x^{*4}) = \sum_{j=0}^\infty V_{x^{*4}}^j({\bf x}^*) \left(\nabla^{*2}\right)^j   \varphi_{k_1^*,k_2^*}({\bf x}^*, x^{*4}),
\label{eq:Pot_CM}
\end{eqnarray}
where $\mu=m/2$ is the reduced mass and the subscript $x^{*4}$ of $V^j$ represents the scheme to define the potential with the relative Euclidean time separation $x^{*4}$.  (Usually, the potential is defined at the equal time, $x^{*4}=0$.)

Using  $\varphi_{k_1^*,k_2^*}({\bf x}^*, x^{*4})= \varphi_{k_1,k_2}({\bf x}, x^{4})$ and \eqref{eq:Pot_CM}, we obtain\cite{Aoki:2019pnq}
\begin{eqnarray}
{1\over 2\mu} \left(\nabla^{*2}+{\bf k}^{*2} \right) \varphi_{k_1,k_2}({\bf x}, x^{4}) 
&=& \sum_{j=0}^\infty V_{\gamma(x^{4}-i{\bf v}\cdot{\bf x}_\parallel)}^j\left({\bf x}_\perp, \gamma({\bf x}_\parallel+i{\bf v} x^4)\right) 
\left(\nabla^{*2}\right)^n 
 \varphi_{k_1,k_2}({\bf x}, x^{4}),
 \label{eq:Pot_Lab}
\end{eqnarray}
 where $\nabla^{*2}= \nabla_\perp^2+\gamma^2(\nabla_\parallel+ i{\bf v} \partial_{x^4})^2 $ and
 ${\bf k}^{*2}=W/4-m^2$.
 
Since ${\bf x^*}_\parallel$ becomes complex for non-zero $x^4$, a meaningful potential is obtained from \eqref{eq:Pot_Lab} only at $x^4=0$. In addition, we take ${\bf x}_\parallel =0$ to extract the equal time scheme potential.
For example, the effective LO potential in the equal time scheme becomes\cite{Aoki:2019pnq}
\begin{eqnarray}
V^{\rm LO}_{x^{*4}=0}({\bf x}^*_\perp={\bf x}_\perp, {\bf x}_\parallel^*=0)&=&
\left.{\left\{ \nabla_\perp^2+\gamma^2(\nabla_\parallel+ i{\bf v} \partial_{x^4})^2 +{\bf k}^{*2}\right\}  \varphi_{k_1,k_2}({\bf x}, x^{4})
\over 2\mu  \varphi_{k_1,k_2}({\bf x}, x^{4})}\right\vert_{x^4=0,{\bf x}_\parallel=0},
\end{eqnarray}
where we set $x^4=0$ and ${\bf x}_\parallel =0$ after  taking derivatives in the right-hand side.

\subsection{Time-dependent method in the laboratory frame}
 For the extraction of the potential, we employ the time-dependent HAL QCD method\cite{HALQCD:2012aa}, which is given in the CM frame as
\begin{eqnarray}
V^{\rm LO}_{x^{*4}}({\bf x}^*)R({\bf x}^*,x^{*4}, X^{*4})&=&
\left({\nabla^{*2}\over 2\mu} -\partial_{X^{*4}}+{1\over 4m} \partial_{X^{*4}}^2\right) 
R({\bf x}^*,x^{*4}, X^{*4}), 
\end{eqnarray}
where $R$ is  the 4-pt function divided by the 2-pt function squared, expressed by
\begin{eqnarray}
R({\bf x}^*,x^{*4}, X^{*4}) &=& \sum_n A_n \varphi_{W_n^*}({\bf x}^*, x^{*4}) e^{-(W_n^* -2m) X^{*4}} + \cdots, \quad
W_n^* := 2\sqrt{{\bf k}_n^{*2} + m^2},
\end{eqnarray}
 $A_n$ is an overlapping of the $n$-th eigenstate to the source operator, and ellipses represent inelastic contributions. 
 
 The time-dependent formula in the laboratory frame is more involved, and is given by\cite{Aoki:2019pnq}
 \begin{eqnarray}
V^{\rm LO}_{x^{*4}=0}({\bf x}^*_\perp={\bf x}_\perp, {\bf x}^*_\parallel=0) &=&
 \left.
{ (L_\perp + L_\parallel + m E ) ({\bf x},x^4,X^4)\over m G({\bf x},x^4,X^4)}\right\vert_{x^4=0,{\bf x}_\parallel=0},
\label{eq:t-depLab}
\end{eqnarray}
where
\begin{eqnarray}
G({\bf x},x^4,X^4) &=& \left[ (\partial_{X^4} -2m)^2-{\bf P}^2\right] R({\bf x},x^4,X^4),\\
E({\bf x},x^4,X^4) &=& \left[ {\partial^2_{X^4}\over 4m} -\partial_{X^4}-{{\bf P}^2\over 4m}\right] G({\bf x},x^4,X^4),\quad
L_\perp({\bf x},x^4,X^4) = \nabla_\perp^2 G({\bf x},x^4,X^4), \\
L_\parallel({\bf x},x^4,X^4) &=& \left[ -(\partial_{X^4} -2m)\nabla_\parallel+i{\bf P}\partial_{x^4}\right]^2 R({\bf x},x^4,X^4).
\end{eqnarray}
Note that  the right-hand side in \eqref{eq:t-depLab} are  written  in terms of quantities  in the laboratory frame only, 
while the potential in the left-hand side is defined in the CM frame. 

\section{Numerical results for the $I=2$ $\pi\pi$ system} 
We apply the formula in the laboratory frame to the $I=2$ $\pi\pi$ system for the demonstration.

\subsection{Simulation details}
We employ the 2+1 flavor ensemble generated by the CP-PACS collaboration\cite{PACS-CS:2008bkb}
on a $32^3\times 64$ lattice with the Iwasaki gauge action at $\beta=1.90$, the non-perturbatively $O(a)$ improved Wilson action at $c_{\rm SW} =1.7150$ and hopping parameters $(\kappa_{ud},\kappa_s)=(0.13700,0.13640)$,
which leads to $a\simeq 0.091$ fm for the lattice spacing and $m_\pi\simeq 700$ MeV for the pion mass.
All correlation functions are evaluated by the one-end trick\cite{McNeile:2006bz} on 399 configurations $\times$ 16 source time locations with the periodic boundary condition in all directions, and their errors are estimated by the jack-knife method with bin size 21.   
A single $Z_4$ noise is used in the one-end trick with dilutions for color, spinor and even/odd spatial indices.
Quark sources are smeared as $q_s({\bf x},t) =\sum_{\bf y} f({\bf x}-{\bf y}) q({\bf y},t)$ with the Coulomb gauge fixing, where the smearing function is given by $f({\bf x}) =A e^{-B \vert{\bf x}\vert}$ with $B=0.3$.
We take $A=1$ for $\vert {\bf x}\vert =0$, $A=1.2$ for $\vert {\bf x}\vert  < {L-1\over 2}$ with $L=32$, or $A=0$ otherwise.
We consider two laboratory frames, the case 1 with ${\bf P}=(0,0,2\pi/L)$, where pions have momenta ${\bf P}$ and {\bf 0}, 
and the case 2 with ${\bf P}=(0,0,4\pi/L)$, where each pion has a momentum ${\bf P}/2$.
In addition, we also consider the CM frame (${\bf P}=(0,0,0)$) for comparisons. 
 
\subsection{$I=2$ $\pi\pi$ Potentials and scattering phase shifts}
\begin{figure}[bth]
\centering
  \includegraphics[angle=0, width=0.49\textwidth]{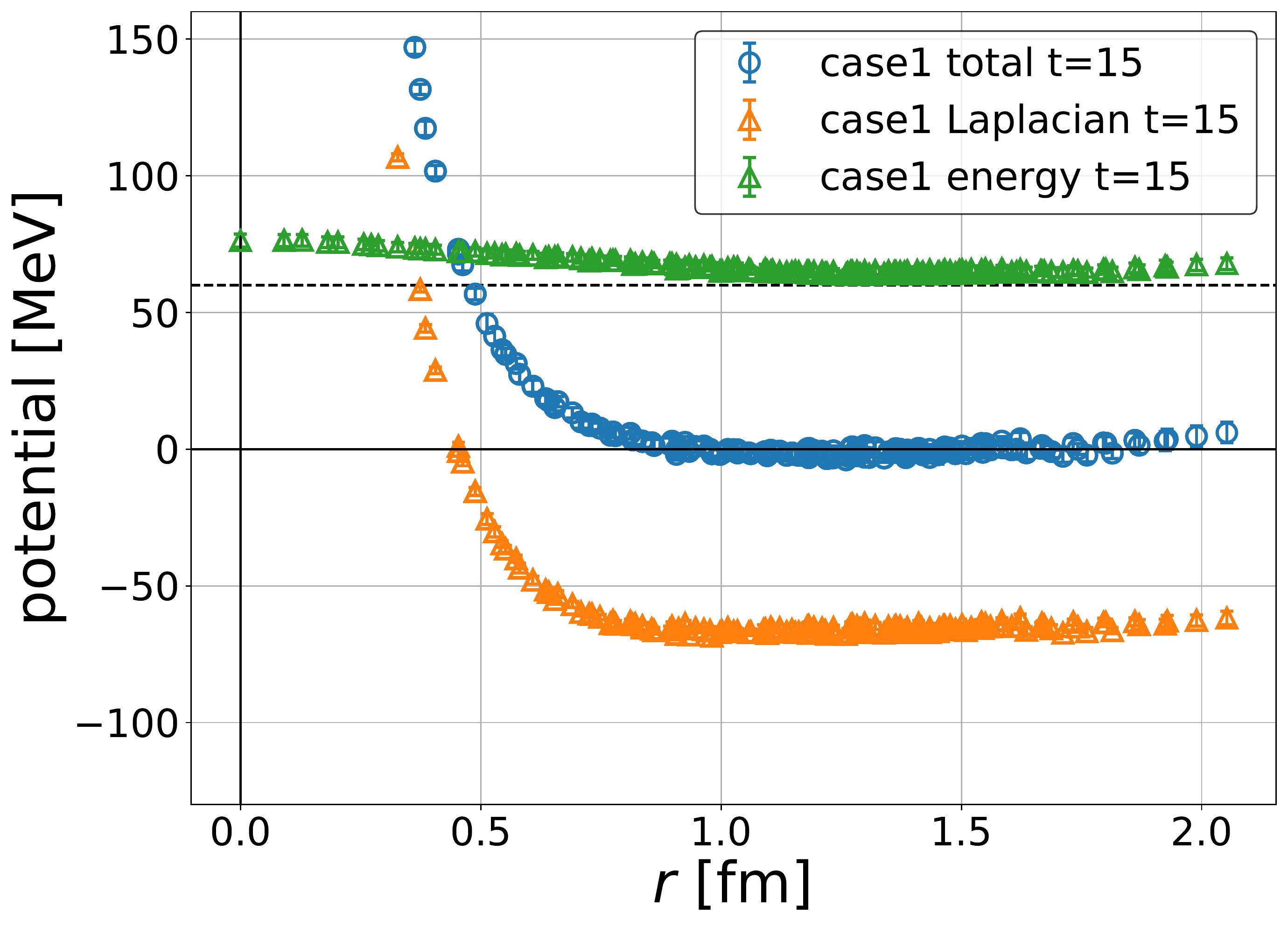}
  \includegraphics[angle=0, width=0.49\textwidth]{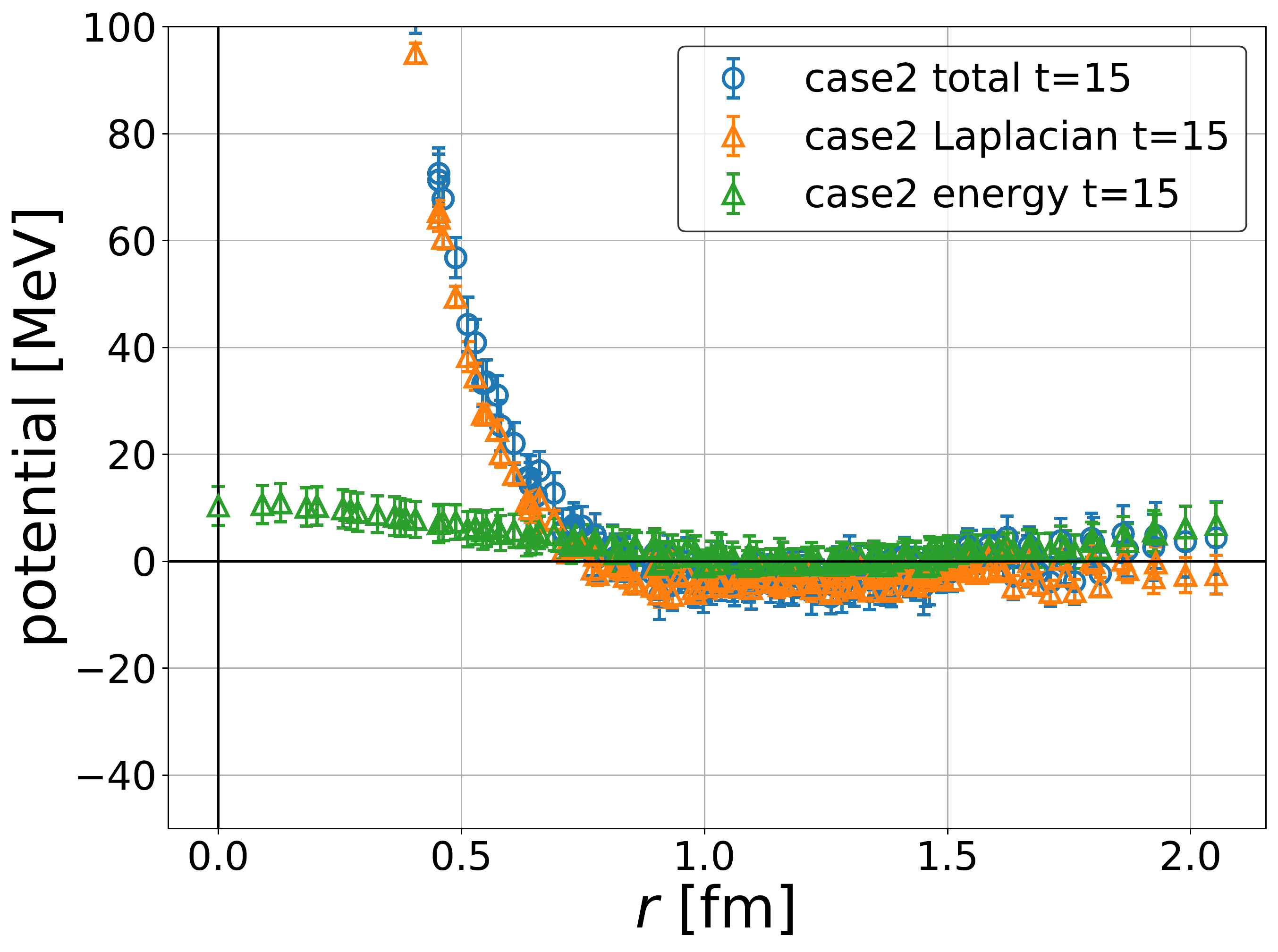} 
 \caption{ Decompositions of the effective LO potential in case 1 (Left) and case 2 (Right), where dotted lines represent expected relative energies in non-interacting cases.
 }
 \label{fig:pot_dcmp}
\end{figure}

Fig.~\ref{fig:pot_dcmp} shows the effective LO potential and its decomposition at $t=X^4=15$ 
for the case 1 (Left) and the case 2 (Right), which is given by
\begin{eqnarray}
V^{\rm LO}_{x^{*4}=0}({\bf x}^*_\perp={\bf x}_\perp, {\bf x}^*_\parallel=0) &=&
\left. \left.
{(L_\perp + L_\parallel )({\bf x},x^4,X^4)\over m G({\bf x},x^4,X^4)}\right\vert_{x^4=0,{\bf x}_\parallel=0}
+ {E({\bf x},x^4,X^4)\over  G({\bf x},x^4,X^4)}\right\vert_{x^4=0,{\bf x}_\parallel=0},
\end{eqnarray}
where the first term represents the Laplacian contribution (orange triangles), the second one the kinetic energy contribution (green triangles), 
and their sum is the effective LO potential (blue circles).  
For the case 1 (Left), the large negative shift from zero of the Laplacian contribution at long distances is compensated by the  large positive shift of the kinetic energy contribution, which is dominated  by the ground state in this laboratory frame (dotted line in the figure for the non-interacting case), so that the total potential approaches to zero at long distances.
This cancellation between the Laplacian and kinetic energy contributions is an evidence for the validity of the time dependent method 
complicated in the laboratory frame. 
For the case 2 (Right), on the other hand, these shifts at long distances are small, as an expected kinetic energy for the ground state is almost zero.  

\begin{figure}[bth]
\centering
  \includegraphics[angle=0, width=0.49\textwidth]{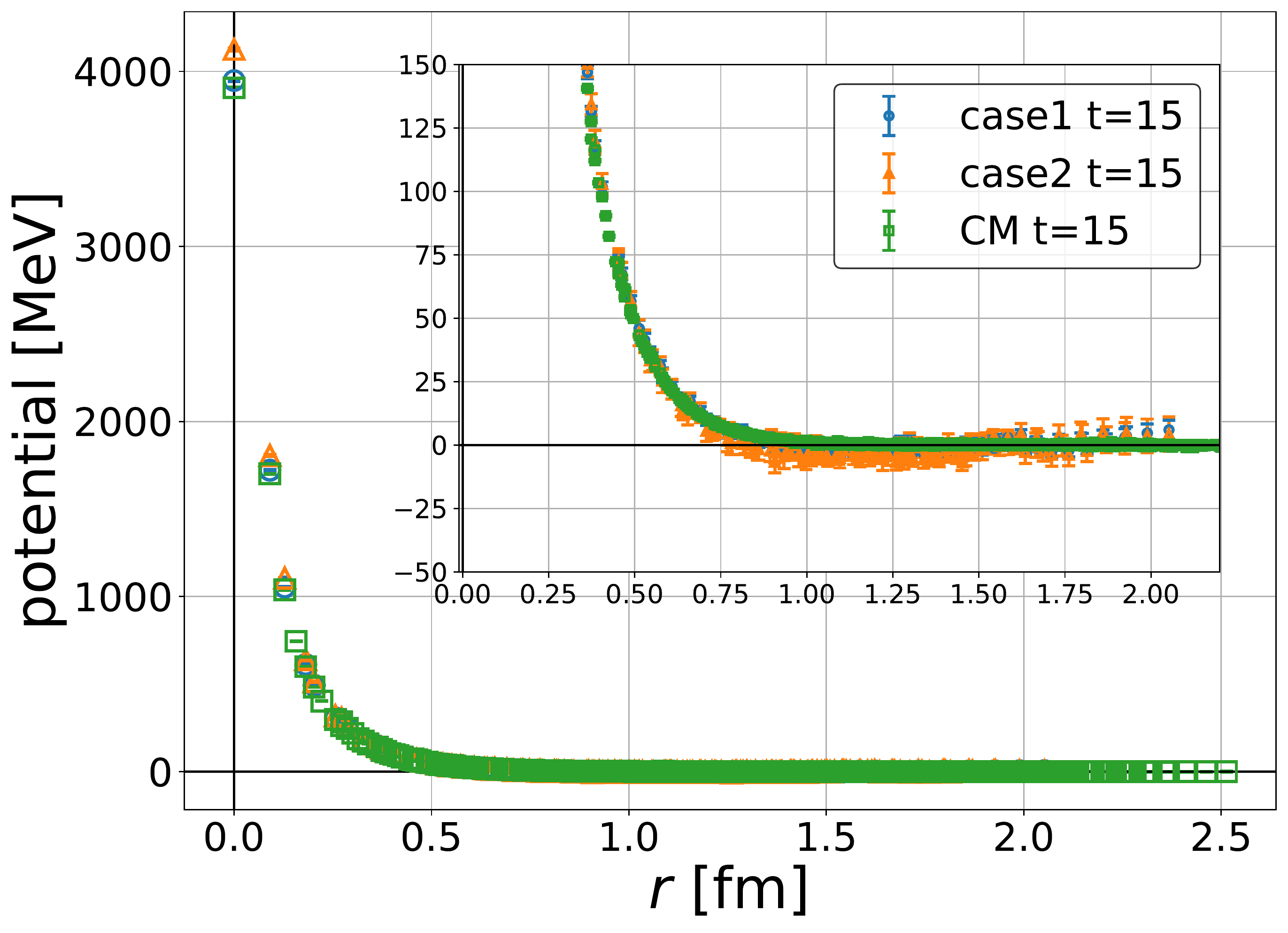}
  \includegraphics[angle=0, width=0.49\textwidth]{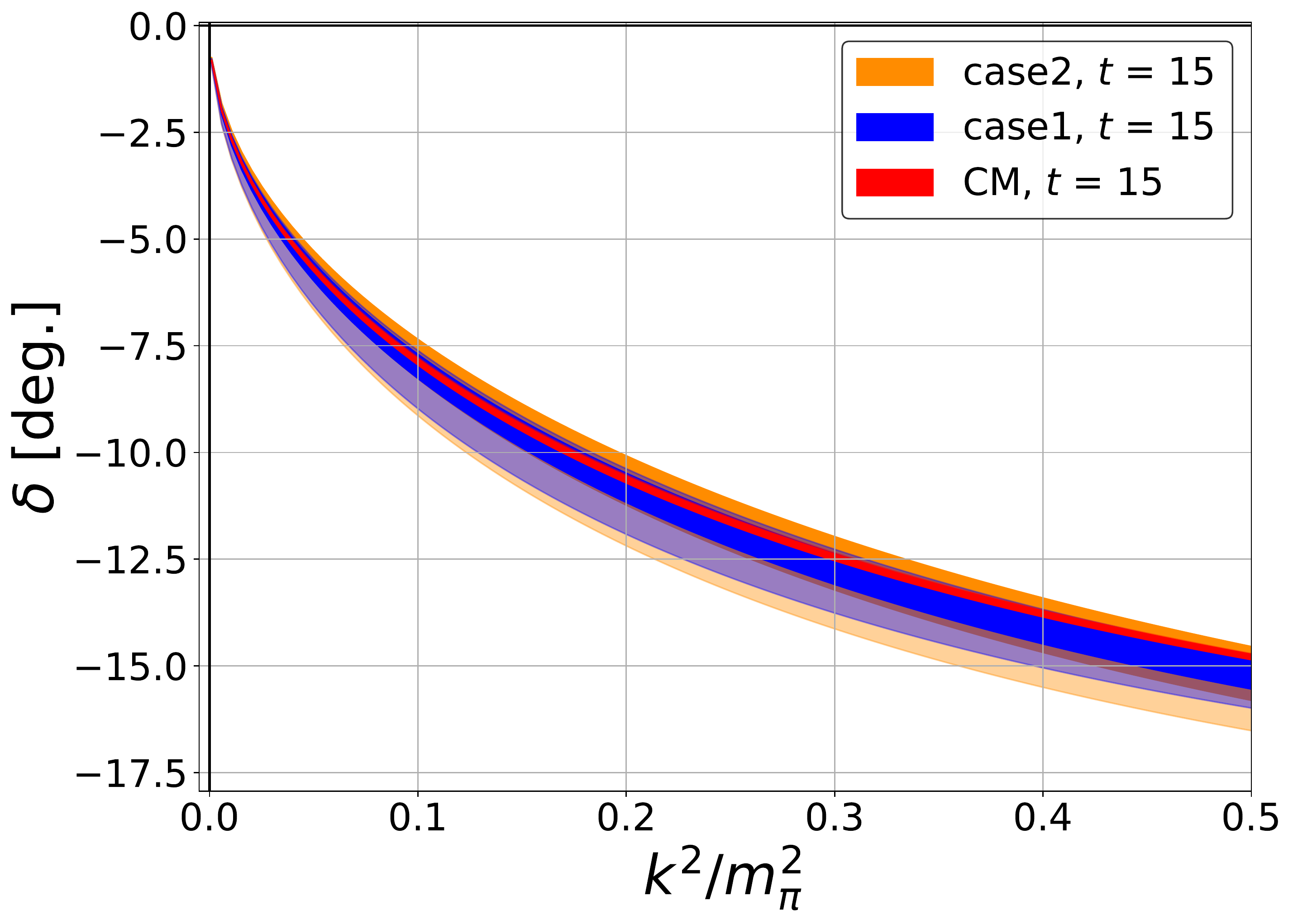} 
 \caption{ (Left) A comparison of the effective LO potentials among the case 1 (blue circles), the case 2 (orange triangles) and the CM (green squares). An inset shows its enlargement. 
 (Right) Scattering phases shifts $\delta_0(k)$ for the $I=2$ $\pi\pi$ system as a function of $k^2/m_\pi^2$ for the case 1 (blue), case 2 (orange) and the CM (red), where dark and light color bands show statistical and systematic errors, respectively.  
 }
 \label{fig:pot_comp}
\end{figure}
In Fig.~\ref{fig:pot_comp} (Left), we compare the effective LO potentials among the case 1 (blue circles), the case 2 (orange triangles) and the CM (green squares), which are almost identical except at short distances  though statistical fluctuations are much larger in laboratory frames. 
Larger statistical fluctuations may be caused by larger total momentum ${\bf P}$, as similar phenomena have been observed 
for single hadron energies with non-zero momenta in lattice QCD. 
Larger fluctuations may be explained partly by larger contaminations to the $L=0$  contribution from higher partial waves in the laboratory frame with $L=2,4,\cdots$, while contaminations come from $L=4,6,\cdots$ in the CM frame.
 The effective LO potentials show the interaction between $I=2$ $\pi\pi$ is repulsive at all distances.

We next calculate scattering phase shift $\delta_0(k)$ for the $I=2$ $\pi\pi$ system, using these three potentials fitted by a sum of 4 Gaussian functions. While central values of scattering phase shifts are obtained form potentials  at $X^4=15$, 
their systematic uncertainties are estimated by differences between $X^4=15\pm 1$.   
Fig.~\ref{fig:pot_comp} (Right) shows scattering phase shifts $\delta_0(k)$ as a function of $k^2/m_\pi^2$ for the case 1 (blue), case 2 (orange) and the CM frame (red), where statistical and systematic errors are given by dark and light color bands, respectively.
As expected from results for potentials, we observe that all three cases give consistent results though both statistical and systematic
errors increases for larger total momentum ${\bf P}$.
Form this agreement, we conclude that  hadron interactions can be investigated by the HAL QCD method using the laboratory frame
in practice, even though systematic as well as statistical errors become larger.

\subsection{Comparisons with the finite volume method}
\begin{figure}[bth]
\centering
  \includegraphics[angle=0, width=0.49\textwidth]{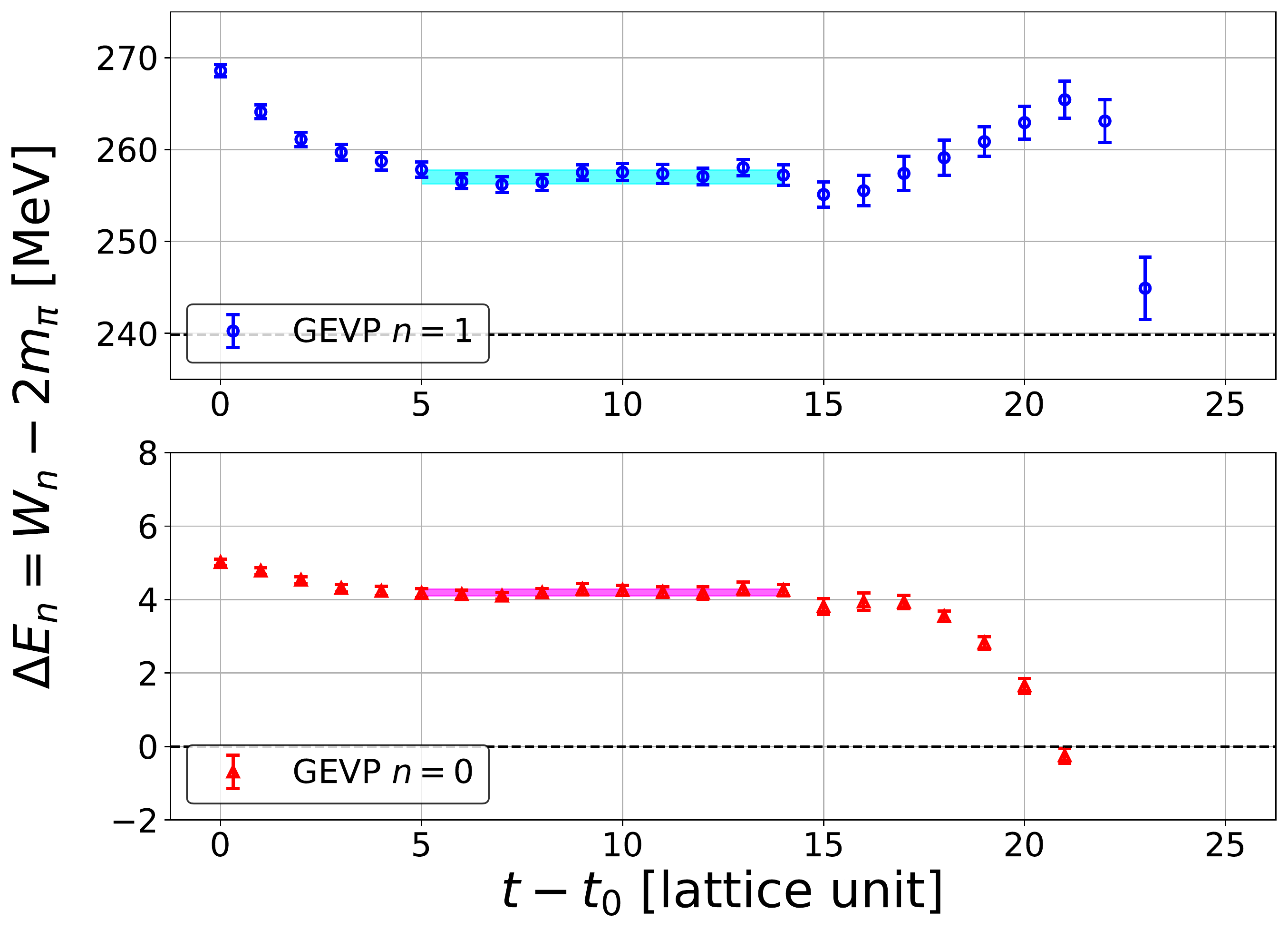}
  \includegraphics[angle=0, width=0.49\textwidth]{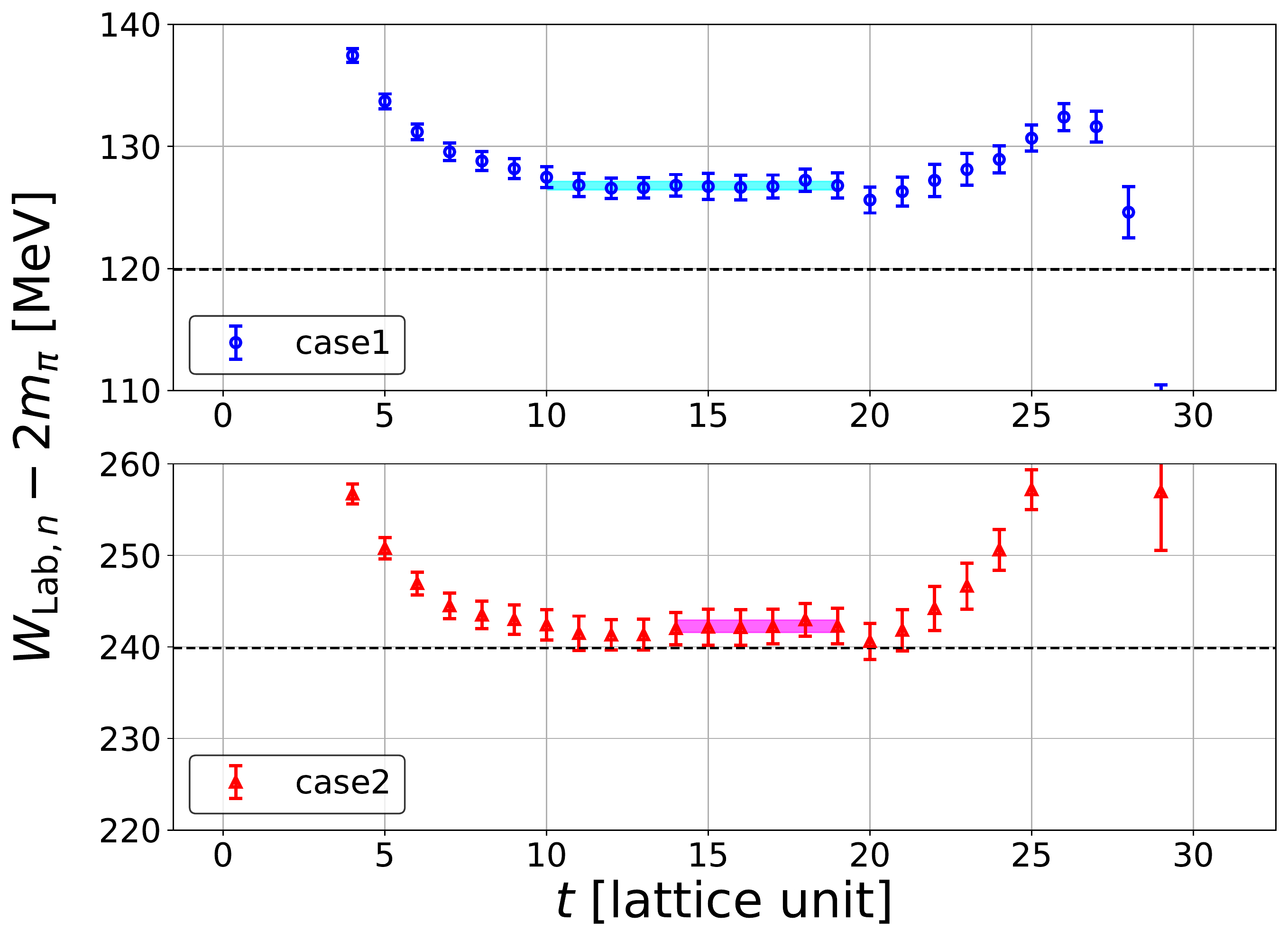} 
 \caption{ (Left) Effective energies of the $I=2$ $\pi\pi$ system, obtained by eigenvalues of the correlation matrix in the CM frame
 for the ground state (red triangles) and the first excited state (blue circles).
 Fits with a single exponential and the corresponding fit ranges are shown by color bands.
 (Right) Effective energies and their single exponential fits for the  ground state of the $I=2$ $\pi\pi$ system in the laboratory frames for the case 1 (blue circles) and the case 2 (red triangles). Dashed and dotted lines are the ground state energies for non-interacting cases. 
 }
 \label{fig:FVspectra}
\end{figure}
We here compare our results to those obtained by the L\"uscher's finite volume method.
We calculate finite volume spectra from correlations functions, two (the ground and the first excited energies) in the CM frame, one each (the ground state) in the case 1 and the case 2.
Fig.~\ref{fig:FVspectra} (Left) shows effective energies of the $I=2$ $\pi\pi$ system for the grand state (red triangles) and the first excited states (blue circles)
obtained by the variational method in the CM frame, while Fig.~\ref{fig:FVspectra} (Right) presents effective energies for the grand state in the case 1 (blue circles) and in the case 2 (red triangles).

\begin{figure}[bth]
\centering
  \includegraphics[angle=0, width=0.49\textwidth]{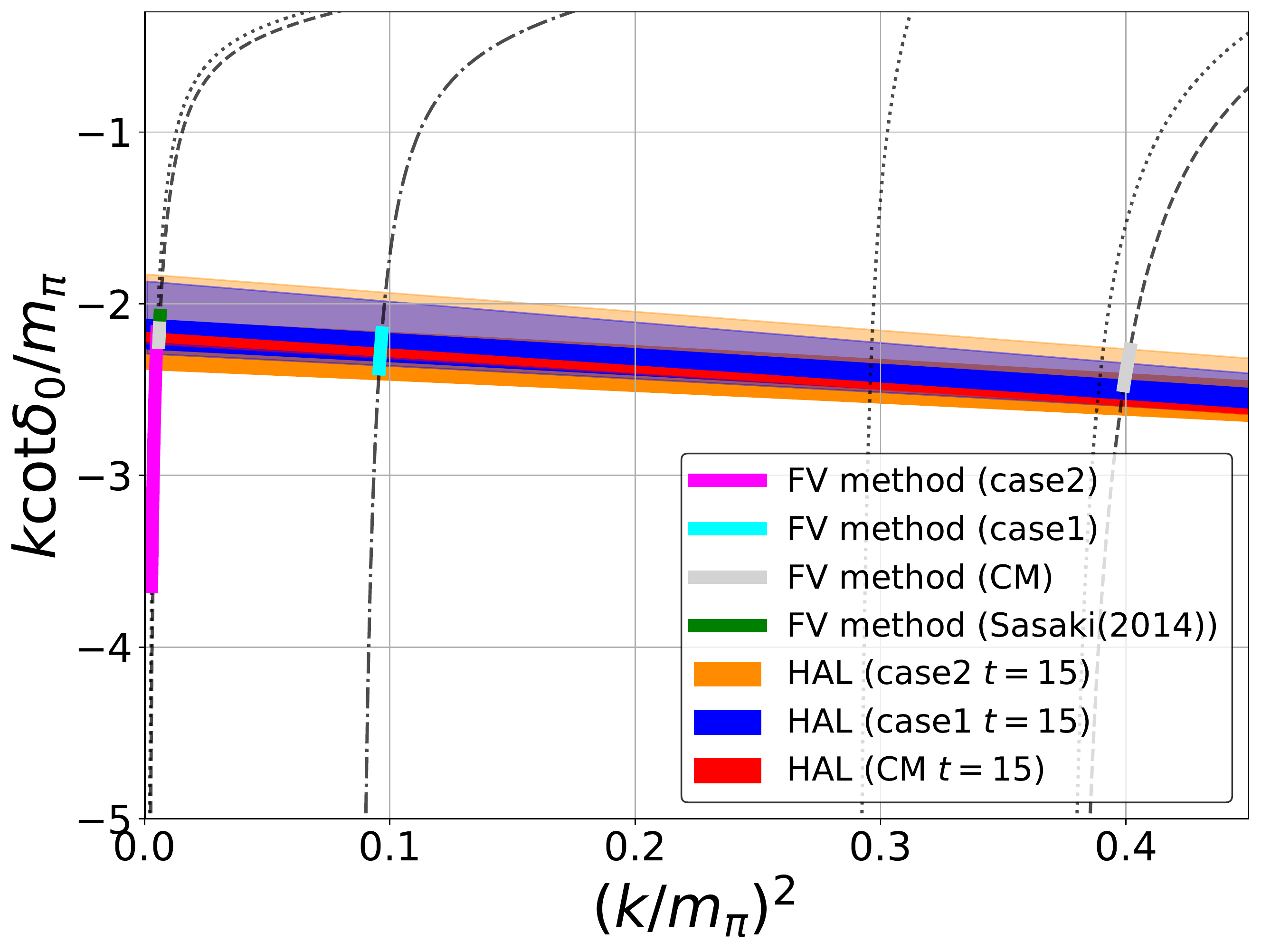}
  \includegraphics[angle=0, width=0.49\textwidth]{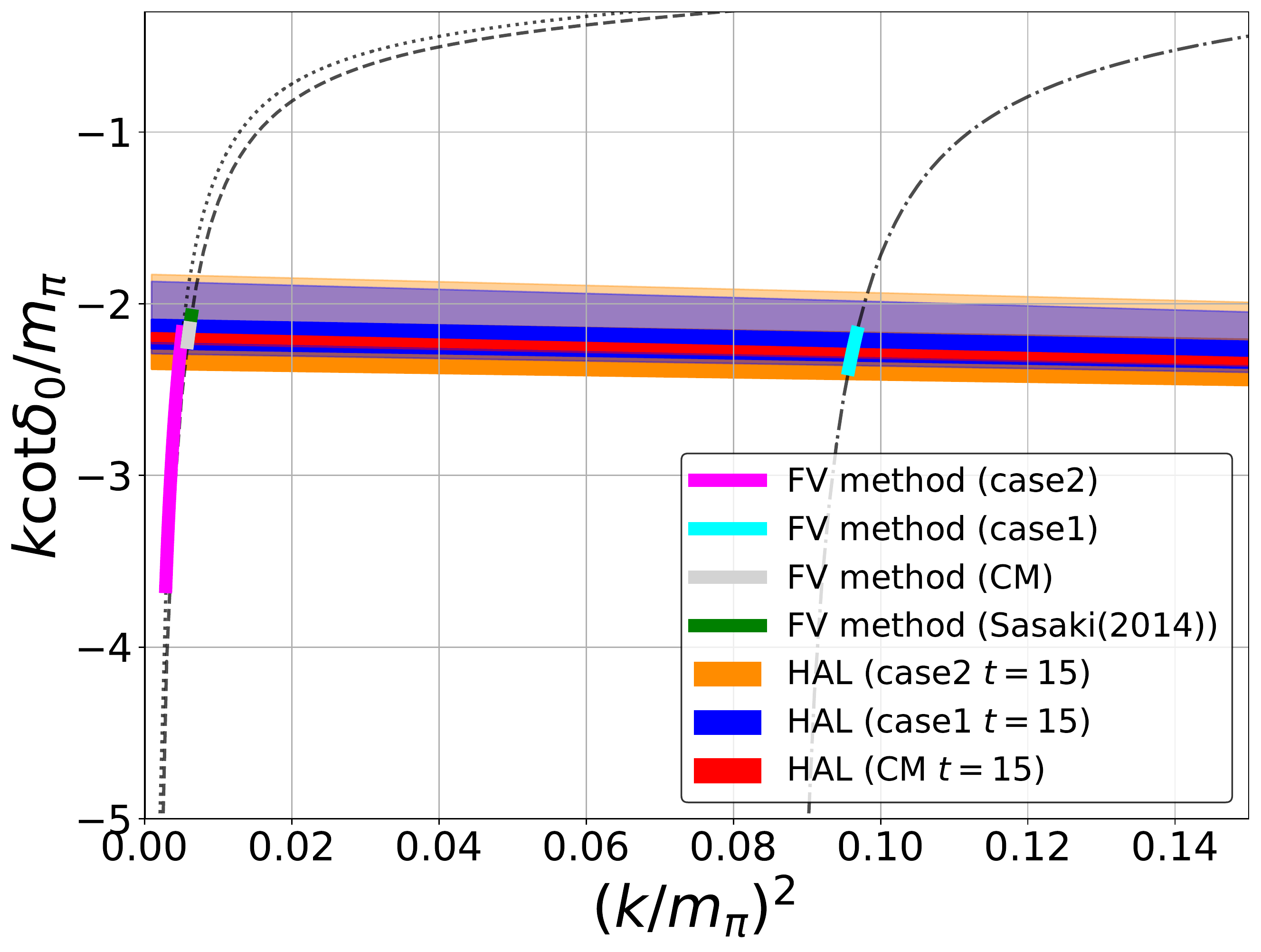} 
 \caption{ (Left) A comparison of $k\cot\delta_0(k)$ with results from the L\"uscher's finite volume  method in the casde 1 (cyan),
 the case 2 (magenta), the CM frame (gray), together with the result from \cite{Sasaki:2013vxa} in the CM frame (green).
 Dash-dot,  dotted, and dashed lines represent constraints by the L\"uscher's formula in  the case 1, the case 2, and the CM, respectively.
 (Right) The enraged one in the low energy region.
  }
 \label{fig:kcotd}
\end{figure}
Fig.\ref{fig:kcotd} summarizes all results in terms of $k\cot\delta_0(k)/m_\pi$ as a function $k^2/m_\pi^2$,
where the HAL QCD results are given by red (CM), blue (case 1) and orange (case) bands representing statistical (dark) and systematic (light) errors, while results from the finite volume spectra are shown by white (CM), cyan (case 1) and magenta (case 2)
bands, together with the CM result from  \cite{Sasaki:2013vxa} by a green band. 
In the figure, constraints by the L\"uscher's finite volume formula are represented by dashed (CM), dash-dot (case 1) and  dotted (case 2) lines. Since finite volume data must lie on these lines, errors of $k$ sometimes result in large errors of $k\cot\delta_0(k)$, as seen in the case 2 (magenta band).

Although errors are larger for both HAL QCD and finite volume data in laboratory frames, 
we confirm that $k\cot\delta_0(k)$ by the HAL QCD potential method are consistent with those by the finite volume method.
These agreements support not only that the HAL QCD method in the laboratory frame works in practice for this system, but also that
the HAL QCD method and the finite volume method agree with each other.
 
\section{Summary}
In this talk, we apply the HAL QCD method with non-zero total momentum to the $I=2$ $\pi\pi$ system.
We show that the method works in practice and gives reasonable results.
This is encouraging since the method opens new possibilities to extract potential of the system with the same quantum number to the vacuum state such as the $I=0$ S-wave $\pi\pi$ system, which contains the $\sigma$ resonance.

\acknowledgments
The authors thank members of the HAL QCD Collaboration for fruitful discussions.
We thank the 
 ILDG/JLDG~\cite{Amagasa:2015zwb} for providing their configurations.
The framework of our numerical code is based on Bridge++ code~\cite{Ueda:2014rya} and its optimized version for the Oakforest-PACS by Dr. I. Kanamori~\cite{Kanamori:2018hwh}.
This work is supported in part
by HPCI System Research Project (hp200108, hp210061),
by the Grant-in-Aid of the Japanese Ministry of Education, Sciences and Technology, Sports and Culture (MEXT) for Scientific Research (Nos.~JP16H03978,  JP18H05236),
by the Grant-in-Aid for the Japan Society for the Promotion of Science (JSPS) Fellows (No. JP20J11502),
by JSPS,
by a priority issue (Elucidation of the fundamental laws and evolution of the universe) to be tackled by using Post ``K" Computer,
by Program for Promoting Researches on the Supercomputer Fugaku (Simulation for basic science: from fundamental laws of particles to creation of nuclei),
and by Joint Institute for Computational Fundamental Science (JICFuS).

\end{document}